# Multiple-Photon Resonance Enabled Quantum Interference in Emission Spectroscopy of $N_2^+$

Xiang Zhang,[1,+] Qi Lu,[1,+] Yalei Zhu,[2] Jing Zhao,[2,#] Rostyslav Danylo,[1,5] Mingwei Lei,[3] Hongbing Jiang[3], Chengyin Wu[3], Zhedong Zhang[4], Aurélien Houard,[5] Vladimir Tikhonchuk,[6,7] André Mysyrowicz [5] Qihuang Gong,[3] Songlin Zhuang,[1] Zengxiu Zhao,[2] Yi Liu,[1,8 *]

[1] *Shanghai Key Lab of Modern Optical System, University of Shanghai for Science and Technology, 516, Jungong Road, 200093 Shanghai, China*
[2] *Department of Physics, National University of Defense Technology, Changsha 410073, China*
[3] *State Key Laboratory for Mesoscopic Physics, School of Physics, Peking University, Beijing 100871, China*
[4] *Department of Physics, City University of Hong Kong, Kowloon, Hong Kong SAR*
[5] *Laboratoire d'Optique Appliquée, ENSTA Paris, Ecole Polytechnique, CNRS, Institut Polytechnique de Paris, 828 Boulevard des Maréchaux, 91762 Palaiseau cedex, France*
[6] *Centre Lasers Intenses et Applications, University of Bordeaux-CNRS-CEA, 351 Cours de la Liberation, 33405 Talence cedex, France*
[7] *ELI-Beamlines, Institute of Physics, Czech Academy of Sciences, 25241 Dolní Břežany, Czech Republic*
[8] *CAS Center for Excellence in Ultra-intense Laser Science, Shanghai, 201800, China*
corresponding author: [#] jzhao@nudt.edu.cn; [*]yi.liu@usst.edu.cn

**Abstract:**

Quantum interference occurs frequently in the interaction of laser radiation with materials, leading to a series of fascinating effects such as lasing without inversion, electromagnetically induced transparency, Fano resonance, *etc*. Such quantum interference effects are mostly enabled by single-photon resonance with transitions in the matter, regardless of how many optical frequencies are involved. Here, we demonstrate quantum interference driven by multiple photons in the emission spectroscopy of nitrogen ions that are resonantly pumped by ultrafast infrared laser pulses. In the spectral domain, Fano resonance is observed in the emission spectrum, where a laser-assisted dynamic Stark effect creates the continuum. In the time domain, the fast-evolving emission is measured, revealing the nature of free-induction decay (FID) arising from quantum radiation and molecular cooperativity. These findings clarify the mechanism of coherent emission of nitrogen ions pumped with MIR pump laser and are likely to be universal. The present work opens a route to explore the important role of quantum interference during the interaction of intense laser pulses with materials near multiple photon resonance.

**Introduction**

High-field physics draws much attention for decades in the study of the materials interacting with intense laser fields. It is responsible for many fascinating effects such as attosecond pulse generation [1], laser acceleration of charged particles [2], precise nano-machining of various materials [3], *etc*. In the framework of strong-field spectroscopy, the light field is often treated classically while the medium is treated quantum mechanically, since the photon number of the driving field is very high so that field fluctuations as typical quantum optical effect is normally neglected [4,5]. Nevertheless, recent works have indicated that quantum coherence and photon statistics of driven and emitted light fields may play a significant role when shining intense femtosecond laser onto gas molecules or semiconductors, creating high harmonics (HHG) [6-9], cavityless lasing of air molecules [10-13], as well as photoionization [14,15]. For instance, Fano interference, a typical quantum optical signature, was reported in the metamaterials producing HHG [16]. This is an indication of optical transition pathway interference imprinted into high-field spectroscopic signals, giving access to the transient dynamics of electrons [17].

Quantum interference is traditionally observed with weak laser field (below threshold field for ionization) resonantly interacting with materials. With laser field close to or more intense than the threshold field for ionization of atoms or molecules, ~ $10^{13}$W/cm$^2$, it was believed that the ionization and plasma formation would destroy the coherence of the system [18-20]. However, this turns out to be not true. In recent studies of strong-field interaction with molecules, dielectric materials and 2D semiconductors, it has been shown that quantum interference and coherence may contribute significantly [6-9]. Quantum coherence is evidenced in the cavity-free lasing of air molecules pumped with intense femtosecond pulses, where the crucial role of the electronic, vibrational and rotational motions of molecules, and their couplings is highlighted [10-13, 21, 22]. Cavity-less lasing of nitrogen molecules in ambient air holds unique potential to generate a virtual lasing source in the sky, where a coherent optical beam is emitted by the air plasma from the sky toward the ground observer [10]. In spite of many efforts [10-13, 20, 23-28], the physical mechanism of cavityless lasing of $N_2^+$ is still controversial. In particular, there is still a debate on the nature of the radiation obtained with different pump laser frequencies, ranging from mid-infrared regime (3.9 μm, 2.6-1.1 μm) to near infrared at 800 nm, and even UV light at 400 nm.

In the present study, we examine the forward 391.4 nm coherent emission corresponding to the transition of nitrogen ions pumped by a MIR wavelength tunable femtosecond pulses in the spectral and temporal domain. In the emission spectrum, a Fano line shape and a broadband continuum are observed under proper conditions. To obtain the temporal profile of the emission, we used a cross-correlation method based on sum-frequency generation of the emission signal with a weak probe pulse. It was found that the 391.4 nm emission presents a pressure-independent monotonous decaying waveform of several picoseconds duration, following fifth/third harmonics generation of a few hundred femtoseconds duration. Our measurements of the emission spectrum reveal the important role of quantum interference between a direct transition channel between the excited and ground state of molecular ion and a continuum induced by the dynamic Stark effect. We attribute the nature of the direct coherent forward emission to free induction decay (FID), enabled by three- or five- photons resonant excitation. An intense narrowband emission is observed in $CO_2$ gas as well, suggesting the universality of quantum interference and FID driven by multiple-photon resonance.

**Experimental setup**

In our experiments, femtosecond laser pulses (12 mJ, 800 nm, 40 fs, 1 kHz) from a commercial laser system were first split in an intense and a weaker pulse, with a dielectric beam splitter. The pulse of 5 mJ energy was used to pump a commercial optical parametric amplifier (OPA) which produced MIR femtosecond pulses tunable from 1.1 to 2.6 μm, with a pulse energy ranging from 30 to 650 μJ. The pulse duration of these MIR pulses was 1.5 times of the incident 800nm pump pulses. They were focused by a convex lens of $f$ = 35 or 50 mm into ambient air or a gas chamber filled with nitrogen gas at different pressures. A gas plasma of ~ 2-5 mm length was formed, which gave rise to intense forward 391.4 and/or 427.8 nm emissions. The emissions after the strong nonlinear interaction with the plasma were collimated by another lens of $f$ = 10 mm. To filter out the strong fundamental MIR pulse and the accompanying white-light emission, the pulse was reflected successively on two dichromatic mirrors, which transmit only spectral components below 550 nm. In a first experiment, the radiation below 550 nm was collected into a fiber tip connected to a spectrometer for spectral measurements. In a second experiment, the 391.4 nm radiation was focused onto a sum-frequency generation (SFG) BBO crystal (cut angle = 44.3°) together with the second weaker pulse at 800 nm with an energy of ~ 50 µJ energy. The SFG signal at 263 nm was recorded as a function of the relative delay between the 319.4 nm radiation and the probe pulse, to resolve the 391.4 nm radiation in the temporal domain.

## Results

Spectrum of the forward emission and the Fano line shape. The experimental setup is presented in Fig. 1, where the femtosecond laser pulse with a tunable central wavelength between 1100 and 2600 nm is focused into a nitrogen gas filled gas cell, with controlled pressure. We present the spectrum of the forward emission for different pump laser wavelengths $\lambda_0$ in Fig. 2. The broadband emission around 370-450 nm in Fig. 2(a), showing a red shift with increasing pump laser wavelength, corresponds to the third harmonic of the MIR pump laser. More importantly, intense narrow lines around 391.4 and 427.8 nm, corresponding to the transition of $B^2\Sigma_u^+$ ($v' = 0$) to the $X^2\Sigma_g^+$ ($v = 0, 1$) states of nitrogen ions, appear for a pump laser wavelength between 1150 and 1320 nm. This agrees with previously reported observations [26]. For a pump laser wavelength tuned between 1530 nm and 1890 nm, a strong emission around 391.4 nm and a relatively weaker one around 358.2 nm (corresponding to transition between $B^2\Sigma_u^+$ ($v' = 1$) to the $X^2\Sigma_g^+$ ($v = 0$)) can be observed, as presented in Fig. 2 (b)-(c). The broadband emission situated around 480-550 nm and 360- 380 nm in Fig. 2 (b) and (c) corresponds to the third and the fifth harmonics of the MIR pump laser, respectively. These results are summarized in Fig.3(a), and compared with simulations, which will be detailed later on.

We notice that the intensity of the narrow line 391.4 nm radiation relative to the intensity of the 3rd/5th harmonics depends on the pump pulse energy, focusing geometry, and gas pressure. By optimizing the pump laser energy and gas pressure, we observed two key features in the spectral domain. For a pump laser wavelength $\lambda_0$ = 1170 nm with a moderate incident energy of 32 µJ, a Fano line shape can be observed clearly in ambient air, as presented in Fig. 4 (a), while for $\lambda_0$ =

1800 nm, we found a continuum between the 5th harmonic and the 391.4 nm radiation, as depicted in Fig. 3(a) and Fig. 4(b).

Time-resolved measurement of the radiation. To gain deeper understanding on the nature of the intense 391.4 nm emission, we performed time-resolved measurements using the cross-correlation technique [28]. Figure 5 presents the experimental measurements of the 391.4 nm signal for the pump laser wavelength of 1950 and 1550 nm. The time-resolved sum frequency signal of the 391.4 nm signal and the weak 800 nm probe pulse are presented in Fig. 5(a) and 5(d). At 1950 nm with a pump pulse energy of $E_{in}$ = 200 μJ, the 391.4 nm emission is not visible so that the fifth harmonic dominates, as shown in Fig. 5 (c). In the temporal domain, the 5th harmonic emission shows a short duration of 250 fs (Fig. 5(b)), as expected from the harmonic generation process. By further increasing the pulse energy, the 391.4 nm emission becomes prominent, and finally dominates in the spectrum (Fig. 5 (c)). Notably, in the temporal domain, following the fifth harmonics, a monotonous decay of emission profile extending to 4 ps is observed. For $\lambda_0$ = 1550 nm whose 3rd and 5th harmonics are far away from the 391.4 nm spectrum range, the emission shows a clean temporal profile during 2-3 ps (10% level), as given in Fig. 5 (e). Moreover, we examined the influence of nitrogen gas pressure, keeping the pump wavelength fixed at 1550 nm. The temporal profile of the emission does not show significant changes within the range of gas pressure from 20 mbar to 70 mbar. Therefore it rules out a superradiance mechanism which should present an emission built-up time within picoseconds and an emission lifetime scaling with the inverse of gas pressure [28-30].

Theoretical model. In support of our measurements, we propose a theoretical model describing the emission spectra of nitrogen molecules subject to a MIR laser pulse, without consideration of the propagation effect. The excitation and emission of nitrogen molecules is treated in the strong-field approximation (SFA) [31]. The laser-induced dynamics of the ions created by the field ionization is described by the two-level model [20, 32], focusing on the coupling between the ionic ground sate $X\,^2\Sigma_g^+$ and the excited state $B\,^2\Sigma_u^+$ both with the vibrational quantum number $v' = v = 0$,

$$i\frac{\partial}{\partial t}\begin{pmatrix} a_X(t) \\ a_B(t) \end{pmatrix} = \begin{pmatrix} \varepsilon_X & -\mu E(t) \\ -\mu E(t) & \varepsilon_B \end{pmatrix}\begin{pmatrix} a_X(t) \\ a_B(t) \end{pmatrix}, \quad (1)$$

where $\varepsilon_{X,B}$ and $a_{X,B}(t)$ denote the energies and the amplitudes of the ionic states $X^2\Sigma_g^+$ and $B^2\Sigma_u^+$, and $\mu \approx 1.5D$ is the transition dipole moment [33]. The electric field is $E(t) = E_0 f(t)\sin(\omega t)$, where $E_0$ and $\omega$ are the peak amplitude and frequency of the laser pulse. For the field envelope $f(t)$, we use a trapezoidal profile with a total duration of sixteen optical cycles and two-cycle linear ramps. For simplicity, we assume that the strong-field ionization creates ions only in the ground state $X^2\Sigma_g^+$. After the laser pulse action, the evolution of the ionic state $B^2\Sigma_u^+$ is described as a decay process with an effective lifetime estimated from the linewidth of the 391.4 nm signal. As a result, the full emission involves harmonic generation by the neutral nitrogen molecules $N_2$ and a coherent emission from the laser-driven nitrogen ions $N_2^+$. The emission spectrum can be obtained from the Fourier transform of the time-dependent polarization,

$$d(t) = d_{XB}(t) + d_n(t), \quad (2)$$

where $d_{XB}(t) = a_X^*(t)a_B(t)\mu$ is the transient polarization between the ionic states X and B states and $d_n(t)$ is the polarization corresponding to the harmonic generation from the neutral ground state [31].

## Discussion

**Fano line shape and intensity-dependent dynamic Stark shift.** The calculated emission spectrum is plotted in Fig. 3 (b). The broadband harmonics and the narrowband 391.4 nm emission show a strong dependence on the pump laser wavelength, in good agreement with experiments. When the 3rd or 5th harmonic spectrum approaches the wavelength of 391.4 nm, the interference between the two transition pathways results in the asymmetric Fano line shape in the spectra [34, 35]. The harmonics emerge during the laser pulse action, providing a broadband background, and the coherent emission at 391.4 nm contributes afterward. The spectral profile is determined by the relative intensity and phase between the two transitions, which are sensitive to the laser intensity. Taking an appropriate ratio of amplitude and phase difference between the 391.4 nm emission and the broadband 5rd harmonic, the Fano feature can be reproduced in Fig. 4 (b).

The model reproduces the observation that the fifth harmonic and the 391.4 nm signal are enhanced at the wavelength around 1800 nm, which deviates from the field-free five-photons resonant wavelength of 1957 nm. This difference can be attributed to the dynamic Stark shift induced by the laser-coupling of the X and B states. When the MIR pulse interacts with the nitrogen ions, the energy levels follow the field adiabatically and return to their proper values after the pulse. The energy shift between these two levels depends on the laser intensity as $\Delta\varepsilon_{XB}(t) \approx \frac{2\mu^2}{\hbar\omega}E^2(t)$. Here µ is the transition dipole moment and $\omega$ the frequency of the laser field. The population at the B state is efficiently enhanced when the driving field couples the X and B states, that is, when the 3rd or 5th harmonic approaches the energy difference between the levels, $\varepsilon_B - \varepsilon_X + \Delta\varepsilon_{XB}$. For example, the laser intensity of $1.3 \times 10^{14}$ W/cm$^2$ corresponds to an optical period averaged energy shift of $\Delta\varepsilon_{XB} \approx$ 0.265 eV. This value matches well 5 times of the energy difference of 0.055 eV between the 1957 nm and the 1800 nm, which corresponds to the resonant coupling to the 5th harmonic at the wavelength of 1800 nm, instead of 1957 nm in the absence of the Stark effect. Agreement with the observed enhancement of emission at this wavelength suggests that it originates from the excitation of B level by the 5th laser harmonic assisted by the dynamic Stark shift and a subsequent free-induction decay of the population at B level to the ground state.

The role of the dynamic Stark shift in the excitation of B level is confirmed by a comparison of the emission spectra around 391.4 nm for two values of laser energy of 70 µJ and 140 µJ in Fig. 6. For lower incident energy of 70 µJ the optimum pump wavelength was observed to be 1830 nm, while it shifts to 1800 nm for $E_{in}$ = 140 µJ. The shift of the optimum wavelength for increasing pump laser intensity is in good agreement with the theoretical predictions, confirming the role of dynamic Stark effect. Since the dynamic Stark effect depends on time, the energy difference between B and X states spans a range between the field-free value $\varepsilon_B - \varepsilon_X$ and the maximum shift $\varepsilon_B - \varepsilon_X + \Delta\varepsilon_{XB}$. As shown in Fig. 3, around $\lambda_0$ = 1800 nm the Stark shift spans the wavelength range from 391.4 nm to 358.2 nm, which explains the excitation of the emission at 358.2 nm observed in the experiment and in the theory.

**Population in the B state versus the pump laser wavelength.** In addition to the strong enhancement of the 391.4 nm emission, the model predicts an oscillatory dependence of emission on the pump laser wavelength, see Fig. 3(b). The origin of this behavior is due to the fact that the population in the B state depends sensitively on the wavelength $\lambda_0$. This prediction is supported by a significant variation of the 391.4 nm signal in Fig. 2(a) and Fig. 2(c) and by observation of the plasma fluorescence, which is directly proportional to the population density in the B state. We present in Fig. 7 the fluorescence signal at 391.4 nm versus the wavelength $\lambda_0$. A clear oscillation with period of ~ 40 nm was observed, which confirms our calculation results.

**Confirmation of the FID effect and its universality.**

The intense narrowband emission, the asymmetrical Fano line shape and the picosecond duration of the emission are the signatures of free induction decay (FID), as reported in the visible and XUV regimes [36-39]. FID is the emission following the coherent excitation of a resonant level that was earlier observed in nuclear magnetic resonance and later in optical domain [36,39]. In the visible and mid-infrared range, both single-photon and multiple-photon resonant excitation induced FIDs have been observed in atoms, molecules and semiconductors [40, 41].

We confirm our hypothesis of free-induction decay with additional evidence in two other aspects. In FID process, there exists no population inversion of the corresponding excited and ground states and no optical gain is expected for an externally injected seeding pulse at the resonant frequency. We confirmed this fact by injecting a seed pulse into the air plasma pumped by a 1900 nm femtosecond pulse: no amplification of the seed pulse was observed, in agreement with the previous reports [25].

The FID effect should be universal for atoms or molecules when an intrinsic optical transition between two electronics levels is resonantly driven by MIR pump laser. This is confirmed in Fig. 8, where we present the emission spectrum of $CO_2$ gas at 100 mbar, inspired by a work by W. Chu *et al* [42]. The pump laser wavelength was 1600 nm and the pulse energy was 80 μJ. The broad emission in the case of a linearly polarized pump pulse corresponds to the 5[th] harmonic of the MIR pump laser, while the narrow peak located around 337 nm is related to the FID transition of $A^2\Pi_u$ ($v = 1$) to $X^2\Pi_g$ ($v' = 0$) of $CO_2^+$.

In summary, we have performed measurements of the intense coherent forward emission at 391.4 nm of nitrogen ions pumped by mid-infrared femtosecond pulses in the spectral and temporal domain. An asymmetrical Fano line shape and a broad continuum bridging the intense 391.4 nm radiation with the 5$^{th}$ harmonic are observed. Time-resolved measurements revealed that the 391.4 nm pulses show a pressure-independent pulse duration of a few picoseconds, which is intrinsically different from the superradiant 391.4 nm emission in case of pumping with 800 nm laser. We attribute this forward 391.4 nm emission to a free induction decay of nitrogen ions, where the B-X transition was resonantly driven by a 3- or 5- photons process. Our theoretical model based on the strong field approximation reproduces well the intense emission at 391.4 nm, the Fano line shape, as well as the spectral continuum with pump laser at wavelength of 1800 nm. This interpretation also explains why an external seed pulse cannot be amplified inside such air plasma. We believe that free induction decay effect should universally exist in atoms or molecules when their intrinsic transitions are resonantly excited by intense femtosecond laser pulse, which is confirmed by observations in $CO_2$.

**Acknowledgement**

The work is supported in part by the National Natural Science Foundation of China (Grants No. 12034013, 91850201, 11904232, 12174011), Innovation Program of Shanghai Municipal Education Commission (Grant No. 2017-01-07-00-07-E00007).

**Figure 1**

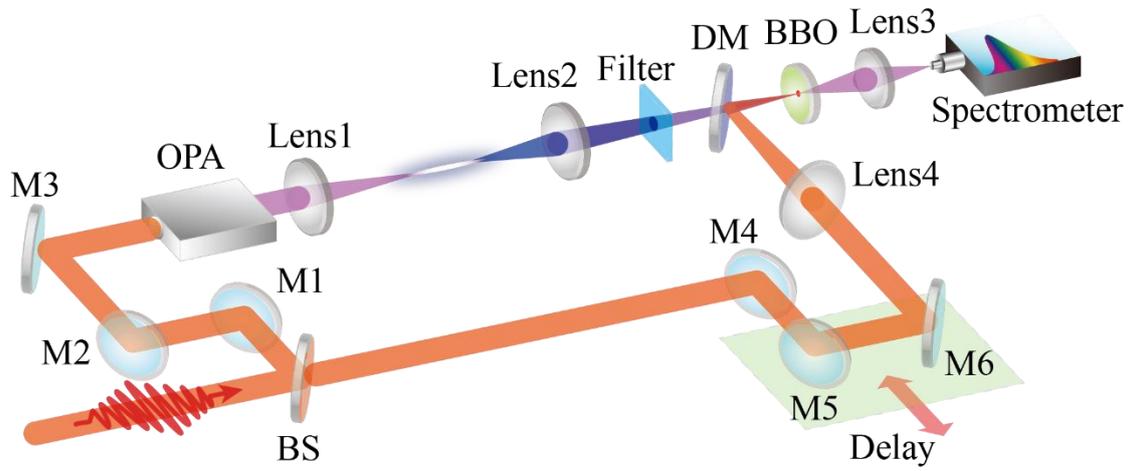

**Fig. 1.** The femtosecond laser pulses pump the Optical Parametric Amplifier (OPA), which provides tunable femtosecond pulses ranging from 1100 mm to 2600 nm. The OPA output pulses pump nitrogen gas and create the plasma. The forward coherent emission from the plasma was filtered and mixed with a weak probe pulse into a BBO crystal for its characterization.

**Figure 2**

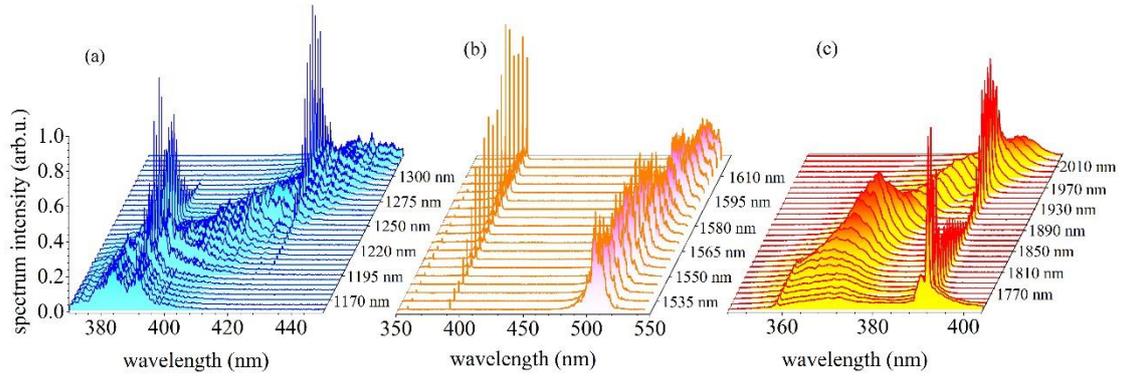

**Fig. 2** Spectrum of the forward emission obtained by different pump laser wavelengths. The wavelength of the pump laser is indicated on the right side of each panel. In (a) and (b) the nitrogen gas pressure was 30 mbar, while in (c) ambient air at 1 bar was used. The pump pulse energy was respectively 70, 90, and 300 μJ for (a)-(c). In (a) and (b), focal lens of $f = 30$ mm was used while in (c) $f = 50$ mm.

**Figure 3**

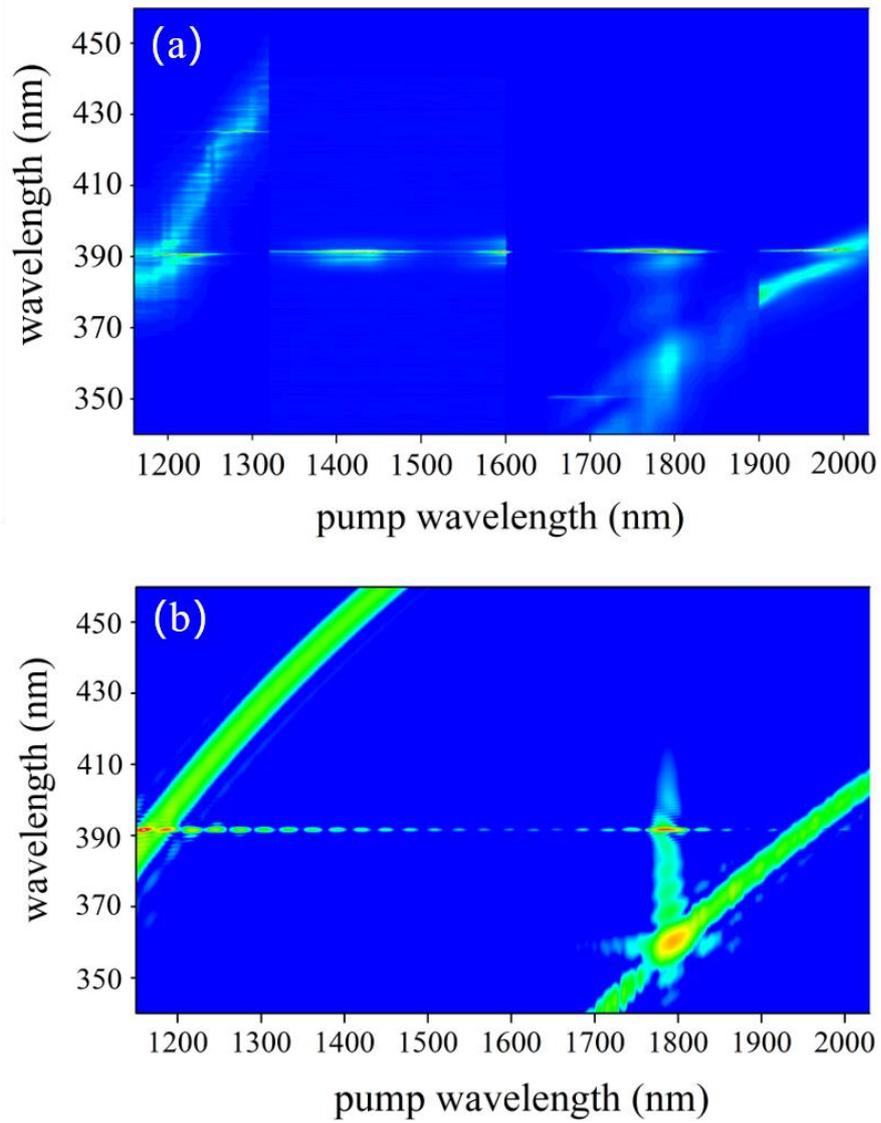

**Fig. 3** Spectrum intensity for different pump laser wavelengths. (a) experimental results, (b) calculated results. The experimental parameters are the same as in Fig. 2. For the simulation, the laser intensity was 1.3 ×10$^{14}$ W/cm$^2$, which corresponds to the enhancement at 1800 nm.

**Figure 4**

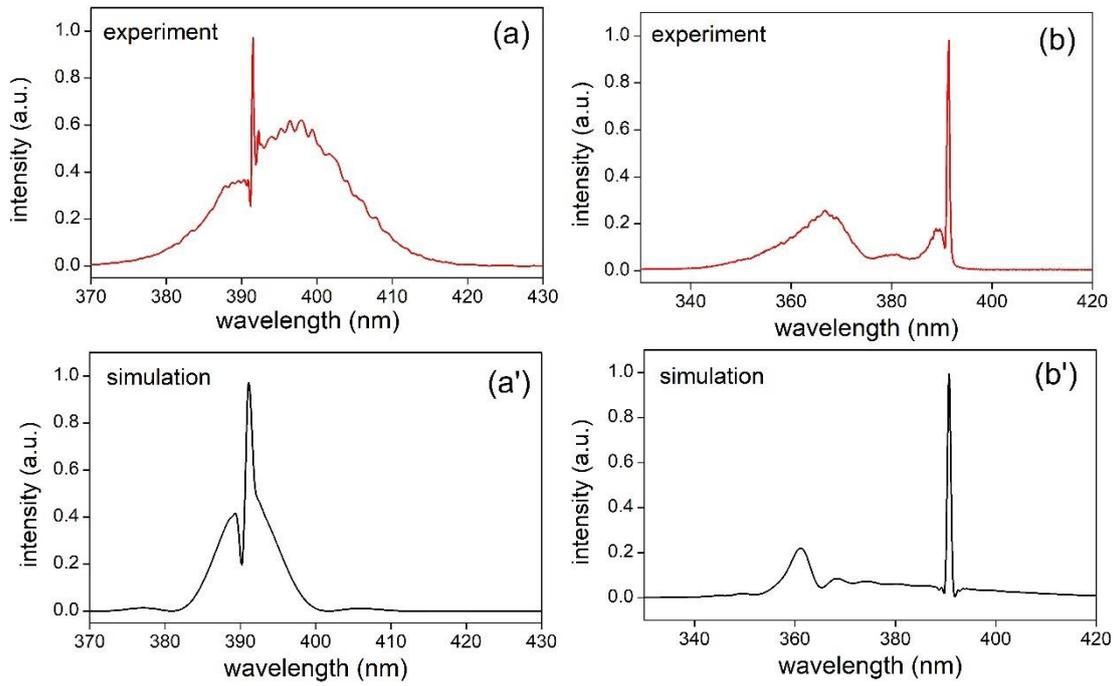

**Fig. 4** Measured and simulated Fano lineshape of the emission spectrum. (a) Spectrum of the forward emission with $\lambda_0$ = 1170 nm, with pulse energy of 32 μJ and 1 bar air used. (b) Spectrum in case of $\lambda_0$=1800 nm. The pump pulse energy was 40 μJ and air is used. (a′) and (b′) corresponding simulation results. The simulation parameters are the same as in Fig. 3.

**Figure 5**

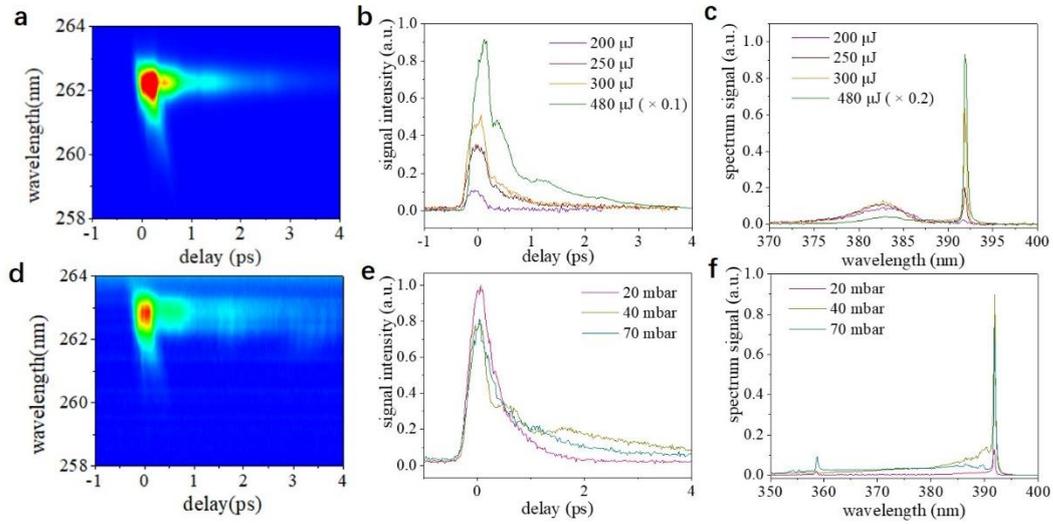

**Fig. 5** Temporal profile and spectrum of the forward 391.4 nm emission. (a) Time resolved measurement of the sum-frequency signal between the 391.4 nm emission and the weak 800nm probe pulse. (b) Temporal profile of the 391.4 nm emission. (c) Spectrum of the forward emission corresponding to (b). In (a) and (c), the temporal profile measurement and spectrum intensity in case of $E_{in}$ = 480 μJ are multiplied by some factors for easier comparison with other energies. The pump laser wavelength for (a)-(c) is 1950 nm. For (d)-(f), the pump laser wavelength is 1550 nm.

**Figure 6**

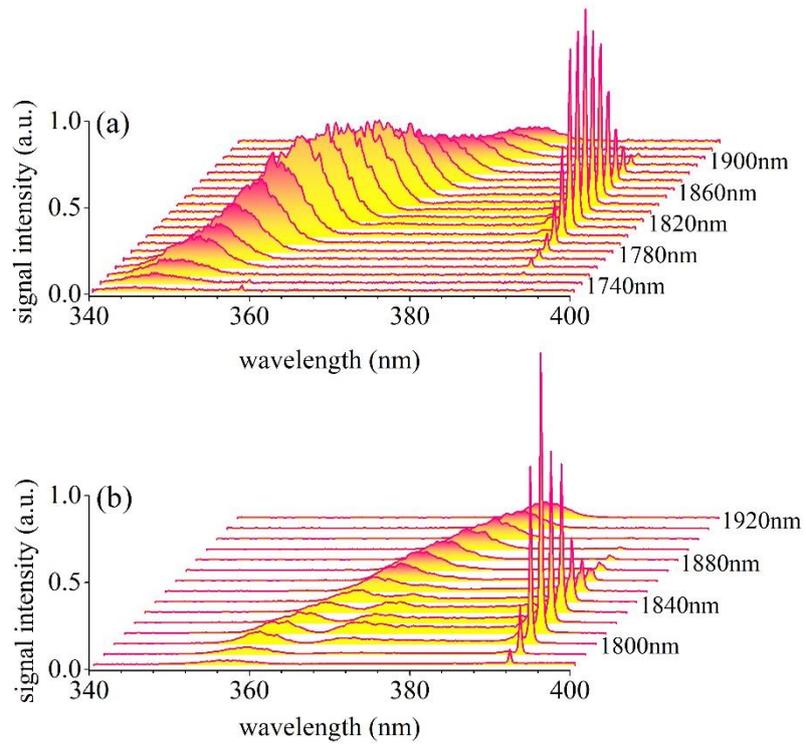

**Fig. 6** Stark effect in the lasing emission. The pump laser energies were 70μJ and 140μJ for (a) and (b). The gas is pure nitrogen of 1 bar. The optimal wavelength for 391.4 nm signal was 1830 nm in (a) and 1810 nm in (b).

**Figure 7**

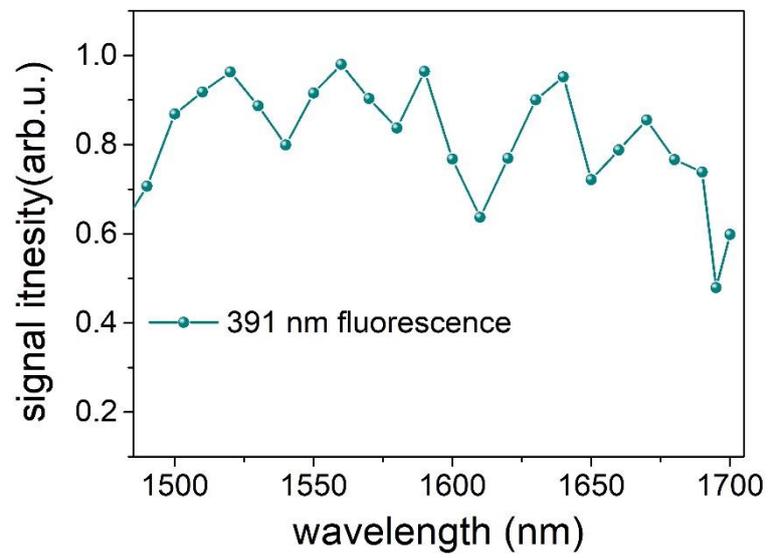

**Fig. 7** Fluorescence at 391.4 nm for different pump laser wavelengths. The energy of the pump laser pulse was 130 μJ and the nitrogen gas pressure was 10 mbar.

**Figure 8**

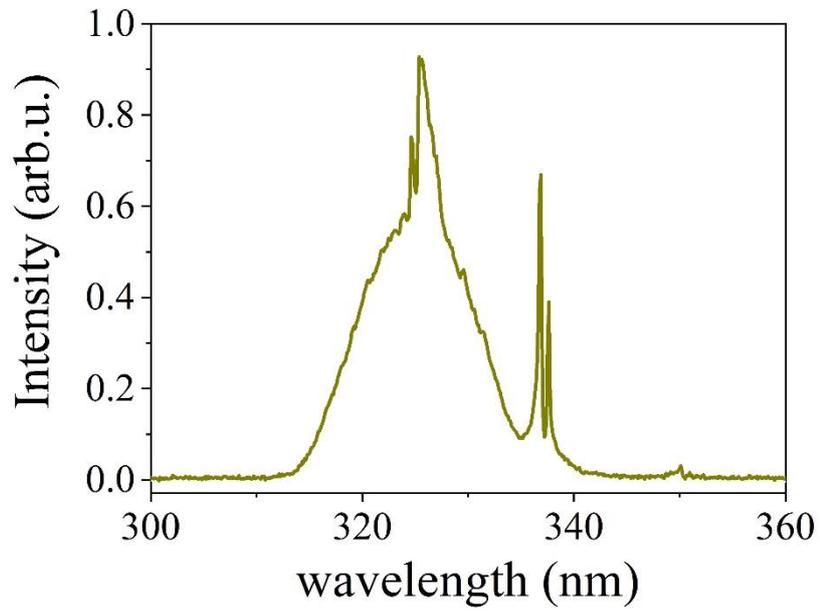

**Fig. 8** Free induction decay signal in $CO_2$. Forward emission spectrum obtained in $CO_2$ gas pumped by MIR pulse at 1600 nm. The pulse energy was 85 µJ and the gas pressure was 100 mbar.